\documentclass[]{article}
\usepackage[preprint]{spconfa4}
\usepackage{graphicx}
\usepackage{babel}

\usepackage{tikz}
\usetikzlibrary{positioning, decorations.pathreplacing, fit, calc}

\usetikzlibrary{external}
\usepackage[subtle, charwidths,leading,indent,lists,bibnotes]{savetrees}

\usepackage[inline]{enumitem}
\setlist{noitemsep}

\usepackage{amsmath}

\usepackage{flushend}

\begin{document}

\title{Training Strategies for {Own~Voice~Reconstruction} in\\ {Hearing Protection Devices} using an In-ear Microphone}

\name{Mattes Ohlenbusch$^{1}$, Christian Rollwage$^{1}$, Simon Doclo$^{1,2}$\thanks{
The Oldenburg Branch for Hearing, Speech and Audio Technology HSA is funded in the program \frqq Vorab\flqq~by the Lower Saxony Ministry of Science and Culture (MWK) and the Volkswagen Foundation for its further development. 
Part of this work was funded by the German Ministry of Science and Education BMBF FK 16SV8811.
This work was partly funded by the Deutsche Forschungsgemeinschaft (DFG, German Research Foundation) - Project ID 352015383 (SFB 1330 C1).}}
\address{
\textsuperscript{1}Fraunhofer Institute for Digital Media Technology IDMT,\\
Oldenburg Branch for Hearing, Speech and Audio Technology HSA, Germany\\
\textsuperscript{2}Department of Medical Physics and Acoustics and Cluster of Excellence Hearing4all, \\
University of Oldenburg, Germany\\
Email: mattes.ohlenbusch@idmt.fraunhofer.de
}


\maketitle

\begin{abstract}
In-ear microphones in hearing protection devices can be utilized to capture the own voice speech of the person wearing the devices in noisy environments. 
Since in-ear recordings of the own voice are typically band-limited, 
an own voice reconstruction system is required to recover clean broadband speech from the in-ear signals. 
However, the availability of speech data for this scenario is typically limited due to device-specific transfer characteristics and the need to collect data from in-situ measurements.
In this paper, we apply a deep learning-based bandwidth-extension system to the own voice reconstruction task and investigate different training strategies in order to overcome the limited availability of training data. Experimental results indicate that the use of simulated training data based on recordings of several talkers in combination with a fine-tuning approach using real data is advantageous compared to directly training on a small real dataset.
\end{abstract}

\begin{keywords}
Own voice reconstruction, in-ear microphone, training strategies, data augmentation, domain adaptation
\end{keywords}

\section{Introduction}
\label{sec:intro}

In noisy working environments, workers often rely on hearing protection devices. Since such devices do not only attenuate external noise, but also hinder direct speech communication, devices enabling radio communication may present an advantage~\cite{nordholm_assistive_2015}. One option for recording the own voice of the person wearing such a device is the use of a microphone placed inside of the occluded ear canal.
However, the in-ear microphone picks up the own voice
at a limited frequency range up to about 2\,kHz
with different transfer characteristics than a close-talking microphone due to occlusion and body-conduction effects and with body-produced noise (e.g., breathing, heartbeats)~\cite{bouserhal_-ear_2019} .
Hence an own voice reconstruction system is required to recover clean broadband speech. 

For this task, it has been proposed in~\cite{bouserhal_-ear_2017} to first use an adaptive filter to reduce noise in the in-ear microphone and then apply a basic bandwidth extension (BWE) system to reconstruct high-frequency content. 
For a similar scenario, in~\cite{liu_bone-conducted_2018} an autoencoder neural network has been applied to directly reconstruct broadband speech from recordings made with a bone-conduction microphone. 
In~\cite{park_speech_2019}, a neural network is utilized to compute time-varying filters used to estimate broadband speech from in-ear recordings.
Recently, in~\cite{yu_time-domain_2020} a multi-modal approach has been investigated using both a bone- and an air-conduction microphone as input signals to a fully convolutional neural network.
In~\cite{liu_multichannel_2020} a fully convolutional neural network approach in the time-domain has been proposed to estimate clean broadband speech using two in-ear microphones. Similarly, in~\cite{li_enabling_2021} it has been proposed to utilize a U-Net architecture to enhance bone-conducted signals in the short-time Fourier transform (STFT) domain.

\begin{figure}[t!]
    \centering
	\includegraphics{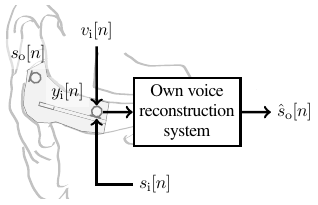}
    \caption{Illustration of the considered hearing protection device and the own voice reconstruction task, aiming at estimating the clean broadband speech signal $s_\mathrm{o}[n]$ from the noisy and band-limited in-ear microphone signal $y_\mathrm{i}[n]$.}
    \label{fig:ovrtask}
\end{figure} 

In this paper, we consider the deep learning-based BWE approach proposed in~\cite{wang_towards_2021}, which reconstructs the high-frequency content using a time-domain U-Net, and adapt it to the own voice reconstruction task, i.e.
not only reconstructing high-frequency content but also compensating for transfer characteristics and reducing body-produced noise.
Since the transfer characteristics are device-specific,
in-ear own voice recordings have to be made in-situ with a talker wearing the device, such that the amount of speech data available for training is typically limited.
We propose to overcome data shortage by simulating artificial recordings for use in data augmentation-based training strategies.
The simulation framework relies on modeling the transfer characteristics between two device microphones using relative transfer functions (RTFs).
We investigate the effects of single- and multi-talker transfer characteristics, the number of RTFs estimated per talker, the influence of body-produced noise, and the usage of real data fine-tuning.
Results indicate that employing a training paradigm adopted from BWE is viable to the own voice reconstruction task. Experimental results show that training on simulated in-ear signals can be used to perform reconstruction on recordings of in-ear signals. In particular, pre-training the proposed system with simulated data and fine-tuning it with real data leads to the largest improvement in terms of objective metrics.

\section{Signal Model} 
\label{sec:sigmodel} 
We consider a scenario where a talker is wearing a hearing protection device equipped with a single microphone located at the inside of the occluded ear (see Fig.~\ref{fig:ovrtask}). 
Since a large component of the own voice speech is transmitted through bone and cartilage~\cite{bouserhal_-ear_2019},
the speech captured by the in-ear microphone exhibits different characteristics than speech captured by a microphone outside of the talkers' body (e.g., a close-talk microphone or a microphone placed at the outer side of the hearing protection device). 
Most prominently, high-frequency components are heavily attenuated, while low-frequency components are amplified. 
It is assumed here that the in-ear microphone does not pick up any external noise from outside of the device, 
but picks up body-produced noise such as breathing sounds.
The considered scenario differs from BWE, since the transfer characteristics may vary based on hearing protection device, ear canal characteristics, and body-produced noise may need to be accounted for. 

Fig.~\ref{fig:ovrtask} illustrates the signal model for the own voice reconstruction task.
In the absence of external noise, the signal $y_\mathrm{i}[n]$ recorded at the in-ear microphone (subscript $\mathrm{i}$ is given by 
\begin{equation}
    y_\mathrm{i}[n] = s_\mathrm{i}[n] + v_\mathrm{i}[n],
\end{equation}
where $n$ denotes the discrete time index, $s_\mathrm{i}[n]$ denotes the own voice speech and $v_\mathrm{i}[n]$ denotes the body-produced noise recorded at the in-ear microphone.
The objective of own voice reconstruction is to estimate a clean broadband speech signal (as it would be captured by a microphone in front of the talkers mouth) from the band-limited and noisy microphone signal $y_\mathrm{i}[n]$. 
Although own voice captured at the outer microphone (subscript $\mathrm{o}$) does not have the same long-term spectrum as speech recorded from a microphone in front of the talker's mouth, we still assume that they are similar such that in this paper we will aim at estimating the own voice captured at the outer microphone.
In this paper, we will use the clean speech signal $s_\mathrm{o}[n]$ captured by a microphone at the outer side of the hearing protection device as the desired speech signal. 
It should be noted that the speech signal $s_\mathrm{o}[n]$ is only used for training and evaluation purposes in this paper, but is  
not available in practice. The speech signals $s_\mathrm{i}[n]$ and $s_\mathrm{o}[n]$ are related to each other by a linear, time-varying transfer characteristic (due to body transmission and mouth movements).

\section{Own Voice Reconstruction System}
\label{sec:ovr_system} 
In this section, we propose a deep learning-based own voice reconstruction system based on the BWE system in~\cite{wang_towards_2021}. In Section~\ref{sec:system}, we describe the used processing framework and network architecture. In Section~\ref{sec:realdata}, we describe a small dataset recorded using the target device. 
To overcome the limited availability of in-ear recordings, in Sections~\ref{sec:simdata} and~\ref{sec:rtfestimation} we propose procedures to simulate in-ear recordings. In Section~\ref{sec:trainingstrat}, we discuss training strategies for the proposed system.

\subsection{Processing framework}
\label{sec:system}
\begin{figure}[t!]
    \centering
    \includegraphics{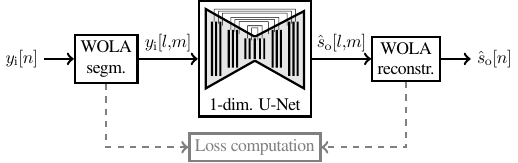} 
    \caption{Overview diagram of the considered own voice reconstruction system.}
    \label{fig:system_structure}
\end{figure}

The overall system is illustrated in Fig.~\ref{fig:system_structure}.
Following~\cite{wang_towards_2021}, it uses a weighted overlap-add (WOLA) framework, where the in-ear microphone signal $y_\mathrm{i}[n]$ is first segmented in segments $y_\mathrm{i}[l,m]$ of length $P$, where $l$ and $m$ denote the segment and segment time index. Each segment $y_\mathrm{i}[l,m]$ is then processed by the U-Net to estimate the clean speech segment $s_\mathrm{o}[l,m]$ at the outer microphone. The estimated segments $\hat{s}_\mathrm{o}[l,m]$ are then reconstructed to obtain 
$\hat{s}_\mathrm{o}[n]$. The WOLA segmentation and reconstruction is carried out using a sqrt-Hann window and an overlap of $\frac{P}{2}$ samples.

In this work, a U-Net architecture as described in~\cite{wang_towards_2021} is utilized. 
This architecture has been used for BWE in~\cite{wang_towards_2021}, but has also previously been applied to noise reduction~\cite{pandey_new_2019, kolbaek_loss_2020}.
Input frames consisting of $P=2048$\,samples are processed in the time domain, at a sampling frequency of $f_s=16\,$kHz.
The input convolutional layer in the encoder part increases the number of channels from 1 to 64,  
and subsequent layers decrease the sampling frequency towards the bottleneck while incrementally increasing the amount of filters. All filters have a length of 11 samples. 
The decoder part of the network after the bottleneck mirrors the encoder part. From each encoding layer, a skip-connection towards the corresponding decoder layer is added. The skip-connections are realized by concatenating the additional channels from the encoder to the output signals of the decoder layer. A parametric rectifier linear unit (PReLU) activation is utilized after each layer except for the last decoder layer, where a linear activation is employed. The U-Net has 10.2M parameters.

\subsection{Real Dataset}
\label{sec:realdata}
A small dataset of own voice recordings was obtained with 14 different talkers (4 female, 10 male)
wearing the closed-vent variant of the commercially available Hearpiece device~\cite{denk_one-size-fits-all_2019} in each ear. 
The microphones used are the in-ear microphone and the outer microphone near the concha. 
Speech was recorded at these microphones while the talkers are reading a German text out loud in a sound-proof listening booth. The talkers were seated during recording, so that body-produced sounds from movement are not expected.
The overall size of the dataset is approximately thirty minutes, which apparently is not sufficiently large to train the proposed system and obtain satisfactory results (see Section~\ref{sec:reconstrresults}). 
In addition, recordings of body-produced noise were gathered from each participant wearning the devices while being silent.

\subsection{Simulated Dataset}
\label{sec:simdata}
Data augmentation strategies may help to overcome limitations imposed by small amounts of available training data~\cite{wen_time_2020}.
We therefore propose to train on a larger speech dataset by simulating data.
In order to easily perform data augmentation,  
we will approximate this transfer characteristic as time-invariant:
\begin{equation}
    \tilde{s}_\mathrm{i}[n] = \tilde{h}[n] * s_\mathrm{o}[n],
    \label{eq:sigmodel_rir}
\end{equation}
where $*$ denotes the convolution operator, $\tilde{s}_\mathrm{i}[n]$ denotes the approximated
own voice speech component at the in-ear microphone and $\tilde{h}[n]$ denotes the relative impulse response (ReIR) between the outer and in-ear microphone, which corresponds to the RTF in time domain. 

In-ear own voice signals are simulated according to the signal model in~\eqref{eq:sigmodel_rir} by convolving broadband speech with ReIRs 
estimated from the real dataset and adding body-produced noise $v_\mathrm{i}[n]$ recorded at the in-ear microphone from the same talkers as in the real dataset, i.e.
\begin{equation}
    \tilde{y}_\mathrm{i}[n] = \tilde{h}[n] * s_\mathrm{o}[n] + \alpha \cdot \tilde{v}_\mathrm{i}[n],
    \label{eq:simsigmodel}
\end{equation}
where the scaling factor $\alpha$ determines the signal-to-noise-ratio (SNR)
of the simulated in-ear signal.

The VCTK corpus containing approximately 44 hours of recordings is used as source material for broadband speech~\cite{veaux_voice_2013}. 
We investigate different options for the ReIR estimates $\tilde{h}[n]$ between the outer microphone and the in-ear microphone, which correspond to RTF estimates in the frequency domain, and the body-produced noise $\tilde{v}_\mathrm{i}[n]$ in \eqref{eq:simsigmodel}:
\begin{itemize}
    \item RTF estimated using recordings from a single talker (1T) vs. RTFs estimated using recordings from all talkers (14T) 
    \item Estimation of a single RTF from a single utterance per talker (s-RTF) vs. estimation of RTFs from a multiple utterances per talker through temporal segmentation (m-RTF)
    \item additive randomly chosen body-produced noise segments, scaled to achieve an SNR randomly varied in [10, 60]\,dB between the in-ear speech and the body-produced noise, included vs. not included  
\end{itemize} 

\subsection{RTF Estimation}
\label{sec:rtfestimation}
First, the own voice recordings are divided into individual utterances using an energy threshold of $-30$\,dB re. maximum peak value per recording for voice activity detection.
Then, STFTs of the microphone signals are computed using a STFT frame size of $N=256$ samples, a Hann window and an overlap of $\frac{N}{2}$ samples between frames.
For each utterance, power spectral density (PSD) estimates are obtained using the Welch method~\cite{welch_use_1967} from the STFTs $Y_\mathrm{i}(k,l)$ and $S_\mathrm{o}(k,l)$ where $k$ and $l$ denote the STFT frequency and frame indices, $L$ denotes the number of frames in an utterance, which varies between utterances, and $\cdot^\dagger$ denotes the complex conjugate:
\begin{align}
    \Phi_\mathrm{i,o}(k) = &\frac{1}{L}\sum_{l=0}^{L-1} Y_\mathrm{i}(k,l)\cdot S_\mathrm{o}^\dagger(k,l) \\ 
    \Phi_\mathrm{o}(k) = & \frac{1}{L}\sum_{l=0}^{L-1} |S_\mathrm{o}(k,l)|^2.
\end{align}
Here, $\Phi_\mathrm{i,o}(k)$ is the cross-PSD between the in-ear and outer microphone signals, $\Phi_\mathrm{o}(k)$ is the PSD of the outer microphone signal.
The relative transfer function $\tilde{H}(k)$ is then estimated as
\begin{equation}
    \tilde{H}(k) = \frac{\Phi_\mathrm{i,o}(k)}{\Phi_\mathrm{o}(k)}
\end{equation}
and the corresponding ReIR $\tilde{h}[n]$ used to generate simulated data is obtained by performing an inverse Fourier transform of the RTF.

However, due to changes in the speech excitation mechanism, it is highly likely that the transfer path changes over time. 
For this reason, speech RTFs are only estimated on individual utterances. In case of the s-RTF option, only the longest utterance is selected from which a single RTF is estimated. In case of the m-RTF option, all utterances with length over 1 second are used to estimate multiple RTFs.

\subsection{Training Strategies}
\label{sec:trainingstrat}
The U-Net (see Section~\ref{sec:system}) is trained using a batch size of $32$ examples per batch, where each example is a single segment of $P$ samples from an utterance randomly chosen from the dataset. Audio input and target signals are normalized to zero mean and unit variance for each recording.
The U-Net is trained using the combined time- and phase constrained magnitude (T-PCM) loss function as proposed for BWE in~\cite{wang_towards_2021}. The loss is computed between the output of the network (estimated clean speech) and either the outer microphone signal in case of the real dataset or the original corpus recording in case of training with the simulated dataset. 
The training is carried out using the Adam optimizer~\cite{kingma_adam_2014} with an initial learning rate of $10^{-4}$, up to a maximum of 100 epochs. 
The learning rate is halved if the validation loss does not improve for 3 epochs, and early stopping is applied after 6 epochs without loss improvement. Dropout with a factor of $0.2$ is performed after each three layers during training.
The real dataset is split based on the device side: recordings and RTF estimates obtained from the left-side device are used in training and validation, whereas the right-side recordings are used as test subset. The training and validation set is further split into proportions of 0.88 and 0.12, respectively.  

Aiming at investigating different training strategies, the U-Net is trained using differently composed datasets. 
First, we investigate suitability only using the real dataset for training. This dataset has the advantage of closely resembling the own voice reconstruction scenario, but has the drawback of limited data availability.

Second, we consider several variants of using the much larger simulated dataset for training.
Since the signal model used to generate the simulated data is only an approximation, differences between simulated and real data may lead to a lower performance than when training with the same amount of data from the real scenario.
Finally, we perform a pre-training of the network on the simulated dataset, and then similarly to~\cite{cheng_transfer_2021} fine-tune only the decoder weights on the real dataset using an initial learning rate of $5\cdot10^{-5}$. It is hypothesized that the encoding features required for own voice reconstruction may be learned from the simulated dataset, and the fine-tuning procedure enables the decoder to better approximate the clean own voice speech at the outer microphone, which is not available during inference.

\section{Experimental Evaluation}
\label{sec:eval}
In this section, we compare the reconstruction performance of the proposed deep learning-based own voice reconstruction system using different training strategies. Additionally, we compare the results to our re-implementation of recently proposed single-channel sinc-dilated fully convolutional network (SDFCN) from~\cite{liu_multichannel_2020}, which is trained on our real multi-talker dataset instead of the single-talker dataset from the original publication. 

\subsection{Evaluation Procedure and Performance Metrics}
For the experimental evaluation, we utilize the test subset of the real dataset. Speech recordings are cut to 10 seconds. 
To assess the own voice reconstruction performance, typical performance metrics used for bandwidth extension and speech enhancement tasks are considered. A metric which is often used to evaluate bandwidth extension systems is the log-spectral distance (LSD)~\cite{gray_distance_1976}.
Additionally, we consider the wideband (WB) setting of the perceptual evaluation of speech quality (PESQ) metric~\cite{international_telecommunications_union_itu_itu-t_2001} and the short-time objective intelligibility index (STOI)~\cite{taal_algorithm_2011}. 
We use the outer microphone signal, assuming to be only own voice speech, as the reference signal for the performance metrics. 
For all measures except LSD, a higher score indicates a better performance. 
LSD is computed on STFT spectra with a frame size of 2048 samples as in~\cite{kuleshov_audio_2017}.

\subsection{Reconstruction Performance}
\label{sec:reconstrresults}
Table~\ref{tab:results_table} shows the experimental results in terms of the considered objective metrics for the unprocessed in-ear microphone signal $y_\mathrm{i}[n]$ and the processed signal $\hat{s}_\mathrm{o}[n]$ using either the baseline SDFCN~\cite{liu_multichannel_2020} or the U-Net for different training strategies.

Here, [R] denotes training on the real dataset, [S] denotes training on a simulated dataset without added body-produced noise, [S+] denotes training on a simulated dataset with added body-produced noise, and [S+R] indicates pre-training on simulated data with added body-produced noise and fine-tuning the encoder on real data. The RTF options are described in Sections~\ref{sec:simdata} and~\ref{sec:rtfestimation}.
\begin{table}[t]
    \centering
    \caption{Mean results for the unprocessed in-ear microphone signal, the baseline SDFCN system and the proposed U-Net system for different training strategies. 
    Best performance is highlighted in bold.}
    \label{tab:results_table}
\small
\begin{tabular}{|l|l|l|c|c|c|c|c|}
\hline
System                             & Data & RTFs used    &  LSD / dB &  PESQ & STOI \\ \hline
unproc.                            & -     &   -         &    2.51 & 1.31  &  0.79\\ \hline
SDFCN                              & [R]   &   -         &    1.53 & 1.47 &  0.74 \\
U-Net                              & [R]   &   -         &    1.48 & 1.64 &  0.73 \\ \hline
U-Net                              & [S]   & 1T, s-RTF   &    1.35 & 1.18 &  0.70 \\
U-Net                              & [S+]  & 1T, s-RTF   &    1.54 & 1.19 &  0.69 \\
U-Net                              & [S+]  & 1T, m-RTF   &    1.51 & 1.26 &  0.74 \\
U-Net                              & [S+]  & 14T, m-RTF  &    1.24 & 1.36 &  0.72 \\ \hline
U-Net                              & [S+R] & 14T, m-RTF  & \textbf{1.05} & \textbf{1.80} & \textbf{0.83}  \\ \hline
\end{tabular}
\end{table}

First, it can be observed that both the SDFCN and the U-Net system trained on the real dataset yield improvements over the unprocessed input signal. 
Compared to the baseline SDFCN, the U-Net with a larger network size yields a higher PESQ score and a lower LSD score, but also a lower STOI score.

When the U-Net is trained with simulated data only, a performance decrease can be observed with respect to using real training data in terms of PESQ, and partly in terms of LSD and STOI. 
For the single-talker, single-RTF training condition, the results in terms of STOI and PESQ are actually worse than for the systems trained with real data. This can probably be attributed to the fact that in this case the U-Net only compensates the static transfer function of a single talker, which does not correspond to real recordings with different and time-varying transfer characteristics. When considering additive body-produced noise and multiple RTFs from one talker in the training dataset, the STOI and PESQ scores slightly improve, but the LSD score degrades compared to the 1T, s-RTF condition. The only training condition where all metrics are improved is the condition where the in-ear microphone signals are simulated using multiple RTFs from multiple talkers. However, it should be realized that even for the 14T, m-RTF training condition the STOI and PESQ scores are still worse than for the systems trained with real data, showing that the assumed signal model in~\eqref{eq:simsigmodel} is probably not realistic enough.
Finally, it can be observed that the training paradigm utilizing the simulated dataset for pre-training and the real dataset for fine-tuning yield large improvements in terms of all metrics compared to both the systems trained on only real data and the systems trained on only simulated data.

Informal listening experiments confirm the signal quality predictions. It should however be noted that while the band-limitation of the in-ear signals appears to be accounted for by the pre-trained and fine-tuned system, there remains an audible difference between the target and the processed signals, probably since the proposed system is unable to account for individual differences in own voice transmission characteristics.

\section{Conclusion}
\label{sec:conclusion} 
In this paper, we have investigated several training approaches for own voice reconstruction from band-limited noisy in-ear microphone recordings. We have proposed a method to simulate in-ear data by utilizing relative transfer functions between an outer and an in-ear microphone of a hearing device. 
Experimental results demonstrate a performance improvement from using simulated data in a pre-training approach. 
For pre-training, the device transfer characteristics seem to be best approximated using a multi-talker, multi-RTF simulation strategy. 
In future work, the influence of individual and device-specific own voice transmission factors and external noise will be further investigated.

\bibliographystyle{IEEEbib-abbrev}
\bibliography{zoterobib.bib}

\end{document}